# Enhanced Habitability on High Obliquity Bodies near the Outer Edge of the Habitable Zone of Sun-like Stars

Christopher M. Colose[1], Anthony D. Del Genio[2], M. J. Way[2,3]

[1]NASA Postdoctoral Program, Goddard Institute for Space Studies, New York, NY 10025
[2]NASA Goddard Institute for Space Studies, 2880 Broadway, New York, NY 10025
[3]Theoretical Astrophysics, Department of Physics and Astronomy, Uppsala University, Uppsala, SE-75120, Sweden

Corresponding author:

Christopher M. Colose
c.m.colose@nasa.gov

ABSTRACT

High obliquity planets represent potentially extreme limits of terrestrial climate, as they exhibit large seasonality, a reversed annual-mean pole-to-equator gradient of stellar heating, and novel cryospheres. A suite of 3-D global climate model simulations with a dynamic ocean is performed with Earthlike atmospheres for low and high obliquity planets with various stellar fluxes, $CO_2$ concentrations, and initial conditions to explore the propensity for high obliquity climates approaching the outer edge of the Habitable Zone to undergo global glaciation. We also simulate planets with thick $CO_2$ or $H_2$ atmospheres, such as those expected to develop near or beyond the outer edge of the Habitable Zone.

    We show that high obliquity planets are hotter than their low obliquity counterparts due to ice-albedo feedbacks for cold climates, and water vapor in warm climates. We suggest that the water vapor greenhouse trapping is greater on high obliquity bodies due to the different dynamical regimes that occur between the two states.

    While equatorial ice-belts are stable at high obliquity in some climate regimes, it is harder to achieve global glaciation than for a low obliquity planet. Temperate polar conditions can be present at high obliquity at forcings for which low obliquity planets would be in a hard snowball state. We suggest the conditions on high obliquity planets are likely to be more favorable for a robust biosphere to develop approaching the outer edge of the HZ. However, the influence of obliquity

diminishes for dense atmospheres, in agreement with calculations from 1-D Energy Balance Models.

# 1. INTRODUCTION

In recognition of the divergent evolution of terrestrial planets in our solar system (Lammer et al., 2018; Del Genio et al., 2019), and more recently compelled by the detection of thousands of extrasolar planets (Udry and Santos, 2007; Borucki et al., 2011), a long lineage of investigations have quantified the range of orbital distances from a star where liquid water can be stable on a planet's surface (e.g., Kasting et al., 1993; Selsis et al., 2007; Kopparapu et al., 2013, 2014). The availability of a surface liquid water reservoir has traditionally been viewed as a minimally necessary condition for the development of an astronomically detectable biosphere (e.g., Kane and Gelino, 2012 for a review).

Near the outer edge of this 'Habitable Zone' (OHZ), where a planet is susceptible to the fate of global glaciation or atmospheric collapse, it is generally assumed that a body with water and operational plate tectonics will exhibit some degree of climate stability by the negative silicate-weathering feedback (Walker et al., 1981). In this paradigm, the climate on planets near the OHZ is kept temperate by a tug of war between low stellar flux and high atmospheric opacity due to $CO_2$ buildup, at least until $CO_2$ reaches a high enough pressure that Rayleigh scattering (increasing planetary albedo) and the onset of $CO_2$ condensation begin to compete with greenhouse warming (Ramirez, 2018). Indeed, the first estimate of the OHZ (Kasting et al., 1993) was to identify the minimum solar flux required to sustain a 273-K surface temperature, after which additional $CO_2$ could not generate further warming.

However, defining a meaningful OHZ boundary is complicated by the orbital configuration of a planet (Spiegel et al., 2009, 2010; Dressing et al., 2010; Armstrong et al., 2014; Linsenmeier et al., 2015) or the possibility of greenhouse warming by gases such as $H_2$ that only condense at much further orbital distances (Pierrehumbert and Gaidos, 2011; Ramirez and Kaltenegger, 2017), even perhaps to the point of keeping rogue planets habitable (Stevenson, 1999). Furthermore, using a nominal ~273 K global-mean temperature to demarcate (un)frozen states is misleading, as discussed in section 3.

Of concern in this paper is the orientation of the rotation axis (obliquity angle) of a planet relative to its plane of orbit. The intensity and spatial distribution of incident stellar flux is one of the fundamental boundary conditions that constrains the global climate of any rocky planet. The obliquity is a governing parameter that determines the time-mean latitudinal profile of top-of-atmosphere (TOA) incoming flux and its seasonal variability.

At low obliquity, the sub-stellar point remains close to the equator throughout the year, while on high obliquity bodies it seasonally migrates poleward of the 'Arctic' or 'Antarctic' circles. For low obliquity planets, the competition between increasing solar zenith angle and increasing length of daylight as latitude increases results in a relatively flat profile of incident flux in summer poleward of about 70° latitude and complete darkness poleward of about the same latitude in the winter hemisphere. For high obliquity planets, incident sunlight increases sharply and monotonically toward the summer pole to values several times higher than on low obliquity planets, while more of the winter hemisphere receives little or no sunlight (Figure 1).

The geometry of low obliquity planets is such that in the annual mean, more energy is received in equatorial regions than at high latitudes; the latitudinal profile of incident radiation becomes monotonically flatter as obliquity increases until a crossover point at ~54° obliquity (Ward, 1974; Dobrovolskis, 2013) when high latitudes receive more energy than low latitudes. Therefore, the baseline climate state prevailing on a high obliquity body is markedly different than a low obliquity one such as Earth.

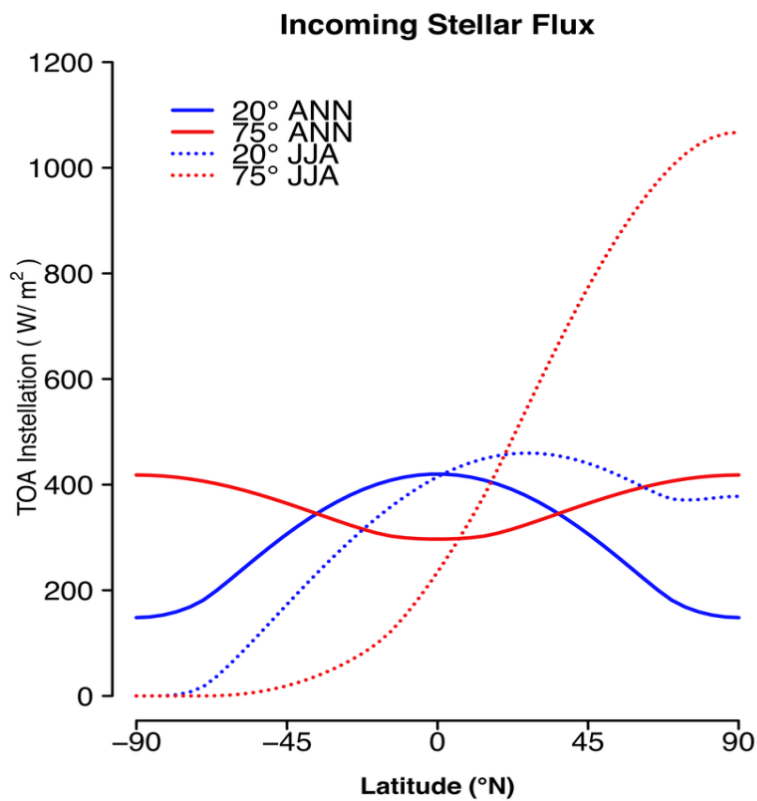

Figure 1: Top-of-atmosphere stellar flux (W m$^{-2}$) vs. latitude for $S_0$=1360 when averaged annually (solid lines) and over June-July-August (dashed lines) for 20° (blue) and 75° (red) obliquity.

For a high obliquity planet operating in a temperature regime that straddles the point of partial glaciation, it is expected that the cryosphere will consist of equatorial ocean ice belts (Spiegel et al., 2009; Rose et al., 2017; Kilic et al., 2018) or low latitude grounded ice sheets over land (Williams, 1993). On such a planet, polar latitudes would experience seasonal or year-round temperate conditions while the equator remained below the freezing point. The distribution of ice cover may also generate bi-stability (multiple possible climate equilibrium points for a given incident stellar flux and atmospheric composition) (Linsenmeier et al., 2015; Rose et al., 2017). Such bi-stability could result in markedly different modern climates on otherwise similar exoplanets with different histories. For instance, a planet with water currently within the habitable zone might or might not be habitable depending on whether it was habitable early in its history or began in a snowball state due to the relative faintness of young host stars (Claire et al., 2012) or inward planetary migration from beyond the ice line (Kuchner, 2003).

It is likely that global mean temperature increases at higher obliquity. This was identified in recent PlaSim simplified general circulation model simulations (Nowajewski et al., 2018) at obliquity sampling between 30° and 90°, and also in the Community Earth System Model (CESM; Kang, 2019). Although the first of these reports "a mere 9 K increase in the global temperature for such large obliquity variations," we suggest that this is a large change and interesting given the globally conservative nature of redistributing incoming energy at the TOA. For perspective, this temperature increase would be expected for ~3 doublings of atmospheric $CO_2$ on Earth and is comparable in magnitude to the largest tropical temperature excursions (relative to modern Earth) during the Cenozoic era (e.g., Frieling et al., 2017; Cramwinckel et al., 2018).

Numerous studies have also found that re-distributing energy in the form of prescribed ocean heat transport from the low to high latitudes increases global temperature (Herweijer et al., 2005; Koll and Abbot, 2013; Rose et al., 2014), a forcing that bears qualitative resemblance to annual-mean obliquity-induced heating (Rencurrel and Rose, 2018). This result is also borne out in simulations isolating the impact of relatively small obliquity variations during Earth's glacial cycles over the last few million years (Mantsis et al., 2011; Erb et al., 2018). Central to these results is the role of the hydrologic cycle, either due to changes in clouds (Rencurrel and Rose, 2018; Kang, 2019) or mid-latitude moistening (Herweijer et al., 2005). We return to this in section 3.2.

The focus of this paper is to explore the role of obliquity on planetary habitability for an Earth-like atmosphere gradually approaching the OHZ of a Sun-like star, and also for planets that have built up multiple bars of either $CO_2$ or $H_2$, at approximately the OHZ or beyond it. We are partially motivated both by results from the intermediate-complexity PlaSim model (Linsenmeier et al., 2015; Kilic et al., 2017), who found that liquid water can remain stable on a planet's surface at orbital distances beyond that which a low obliquity planet can support, and also by energy balance model calculations suggesting that the importance of obliquity should diminish in dense, multi-bar atmospheres (Williams and Kasting, 1997). To our knowledge, this is the first study to present results for multi-bar atmospheres at different obliquities in a 3-D GCM framework, especially in which both the dynamics of the atmosphere and ocean are simulated. Furthermore, this contribution examines climate bi-stability at high obliquity using an independent model from the previous literature, and we build upon the discussion by elaborating on mechanisms for relative global warmth on high obliquity bodies.

## 2. METHODS

To explore the impact of obliquity on habitability near the OHZ, we employ the NASA Goddard Institute for Space Studies Resolving Orbital and Climate Keys of Earth and Extraterrestrial Environments with Dynamics (ROCKE-3D) GCM (described in Way et al., 2017) as our laboratory to perform a suite of simulations at different obliquities, stellar fluxes, and atmospheric compositions.

Most simulations in this paper (except as described in section 3.3) are performed with a dynamic ocean that is 900 m deep. Since the constraints on global climate are fundamentally anchored in the planetary energy budget, scientific study at the intersection between climate and habitability has been usefully engaged with by radiative-convective models (Kasting et al., 1993) or with 1-D latitude-dependent energy balance models (e.g, Williams and Kasting, 1997; Spiegel et al., 2009; Dressing et al., 2010). More recently, emphasis has been given to 3-D experiments employing a thermodynamic slab ocean (i.e., an ocean that exchanges energy and water with the atmosphere but has no ocean currents to transport heat horizontally or sequester heat at depth) underlying an atmospheric general circulation model (GCM, e.g., Williams and Pollard, 2003 Wang et al., 2014; Linsenmeier et al., 2015; Bolmont et al., 2016; Kopparapu et al., 2016, 2017; Popp et al., 2016; Shields et al., 2016; Boutle et al., 2017; Kilic et al., 2017; Wolf, 2017; Wolf et al., 2017; Bin et al., 2018; Haqq-Misra et al., 2018). Fewer planetary studies have performed calculations with an interactive dynamic ocean (e.g., Yang et al., 2013, 2014, 2019; Hu and Yang, 2014; Way and Georgakarakos, 2017; Kilic et al., 2018; Way et al. 2018; Del Genio et al. 2019). Despite the computational convenience of slab ocean simulations, a self-consistent representation of ocean circulation is optimal, particularly for considerations of the OHZ where the latitudinal extent of sea ice is a primary concern. Indeed, it is consistent across a hierarchy of modeling studies that the sea ice edge is shackled to ocean heat transport and the pattern of surface fluxes consistent with the divergence of that transport (e.g., Seager et al., 2002; Bitz, 2005; Rose and Marshall, 2009), with large implications for planetary climates cold enough to develop sea ice (Way et al., 2018).

For simplicity, all runs are conducted in aquaplanet configuration (i.e., no continents) with Earth's gravity, and sidereal rotation and orbital period. The current ROCKE-3D model requires a land point at the south pole, and we added a land point at the north pole for symmetry. These are single grid cells with zero height relative to sea level. Eccentricity is set to zero for all simulations. The ROCKE-3D runs are performed at horizontal resolution of 4°x5° for the atmosphere and ocean, and 40 atmospheric layers extending to 0.1 hPa. The radiation code used is the Suite of Community Radiative Transfer (SOCRATES) radiation scheme (Edwards, 1996; Edwards and Slingo, 1996; see also the description in Way et al., 2017). Spectral resolution is not fixed in SOCRATES; the code is designed for flexibility in choosing bands to optimize simulation time and suitable accuracy for various atmosphere-star combinations. Spectrally dependent data are stored in spectral files that need to be generated once. For all runs we use a Sun stellar spectrum and for $CO_2$ concentrations up to 5,000 ppm, the GA7.0 configuration of SOCRATES is used that features 9 longwave and 6 shortwave spectral bands. For higher $CO_2$ concentrations, there are 17 longwave and 35 shortwave bands in order to capture additional $CO_2$ absorption features that affect radiative heating rates such that the GA7.0 configuration is not sufficiently accurate. $CO_2$-$CO_2$ collision-induced absorption, $CO_2$ sub-Lorentzian line wings, and $CO_2$ self-broadening is included (see discussion in Del Genio et al., 2019 for details).

We perform all simulations at either 20° or 75° obliquity. For each obliquity, we consider stellar constants of 1360, 1075, and 950 W m$^{-2}$ (hereafter, $S_0$=1360, 1075, or 950; the first of these is the modern Earth incident flux). All simulations use a 1 bar atmosphere with $CO_2$ concentrations of 0; 50; 100; 500; 1,000; 5,000; 10,000 ppm; 5%; and 10% $CO_2$. The remaining non-condensable fraction of atmosphere is $N_2$. There is no ozone and therefore no stratospheric temperature inversion.

The computational expense of using coupled atmosphere-ocean experiments precludes finely discretized parameter sampling. We note that the particular values chosen for the stellar constants (see also below for the multi-bar experiments) were selected from a small set of preliminary runs with the goal of identifying radiative forcing regimes that span the range where completely ice-free or snowball conditions occur on planets for both obliquities or where different behavior can occur between them.

In addition to considering incoming stellar flux and atmospheric composition, we also start experiments either with "warm start" conditions, in which the simulation begins with open ocean conditions (above freezing) everywhere, or "cold start" conditions, in which the experiment begins in a complete snowball state with a global-mean temperature of about -50°C.

In section 3.4, we present results for multi-bar atmospheres near or beyond the OHZ. In the first set of experiments, we perform simulations of a 5 bar atmosphere with 80% $CO_2$, 1 % $CH_4$, and the remainder $N_2$ at stellar constants of 500, 550, and 600 W m$^{-2}$. These runs use 16 longwave spectral bands and 46 shortwave bands. We also show 10 bar experiments with a mixture of $N_2$ and 20% $H_2$ (15 longwave and 46 shortwave bands are used) at stellar constants of 125, 150, 175, and 200 W m$^{-2}$, beyond the traditionally defined OHZ. $N_2$-$N_2$, $N_2$-$H_2$, $H_2$-$H_2$ collision-induced absorption is supported. We note that in our solar system, the classical maximum $CO_2$ greenhouse based OHZ occurs at ~1.67 AU (or a $S_0$ of ~490 W m$^{-2}$) at a $CO_2$ pressure of approximately 8 bars (Ramirez, 2018).

Because the seasonal cycle is a fast timescale phenomenon, both the heat capacity of the coupled system and the length of the orbital year will be important in setting the range of conditions that a given location experiences over one orbit. Since our multi-bar atmosphere experiments are meant to make contact with conditions that may be expected for planets near the OHZ, we conducted experiments both with Earth's orbital period and a separate set of runs in which the orbital period, P, is increased to be consistent with Kepler's law (see e.g., Kopparapu et al., 2016) for an orbit in which a planet receives the chosen stellar flux around the Sun. That is:

$$P[days] = 365 \cdot \left[\frac{1360}{S_0}\right]^{0.75} \quad (1)$$

For diagnostics purposes in ROCKE-3D, a longer orbital period corresponds to longer "months" as the annual cycle remains divided into 12 "months" that can be much longer or shorter than 30 days (Way et al., 2017, 2018). In the results to follow, we present both annual means and June-July-August (JJA) seasonal means.

## 3. RESULTS

3.1. Temperature and Sea Ice Patterns

Figure 2 shows the latitudinal profile of surface air temperature at 20° and 75° obliquity (shown for $S_0$=1360 and 500 ppm $CO_2$). At 75°, the equatorial latitudes cool relative to the low obliquity experiment by ~10°C and become the coldest part of the planet throughout the year, even colder than the high latitudes of the winter hemisphere. The low latitude cooling is more than offset by warming poleward of 30° latitude, and global-mean annual temperatures are hotter at 75° obliquity by approximately 7.2°C. Furthermore, despite very little hemispheric exposure to stellar radiation, temperatures in the winter hemisphere at 75° obliquity remain higher than those of their 20° obliquity counterpart. That is, increasing obliquity enhances the seasonal temperature range as expected, but does not do so by virtue of increasing summer temperatures and decreasing winter temperatures; rather, seasonality is enhanced while both summer and winter temperatures remain at a much warmer baseline relative to the 20° regime (Figure 2). We discuss mechanisms for increased global temperature as obliquity increases in the following section.

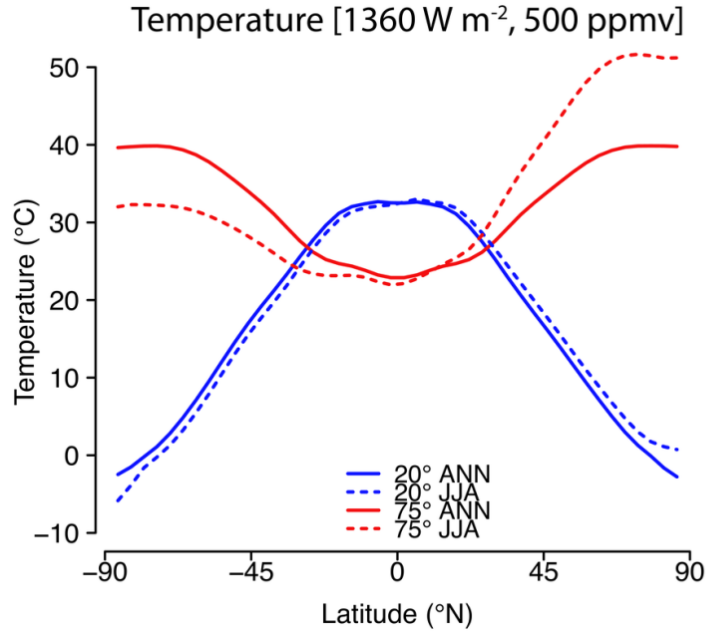

Figure 2: Annual (solid) and JJA (dashed) temperature (°C) for $S_0$=1350 and 500 ppm $CO_2$ at 20° (blue) and 75° (red) obliquity.

Figure 3 presents the dependence of global mean surface temperature on $CO_2$ concentration for the three different values of incident stellar flux (panels a-c), for the low-obliquity and high-obliquity cases (blue and red points, respectively), and using either the warm start or cold start initialization (square and triangle data points, respectively). Figure 4 shows the analogous dependence of global sea ice cover.

Figure 3 shows that the increase in global-mean temperature with obliquity is robust across the entire parameter space of stellar flux and $CO_2$ concentration but that the magnitude of the increase can vary considerably depending on the conditions for each experiment. At $S_0$=1360 (for a given $CO_2$ concentration), global-mean temperatures at 75° obliquity are 7-9 °C higher than at 20° when both planets have no sea ice, and 35-40°C hotter when the low obliquity version exhibits near planetary-scale coverage of sea ice but when the high obliquity version does not.

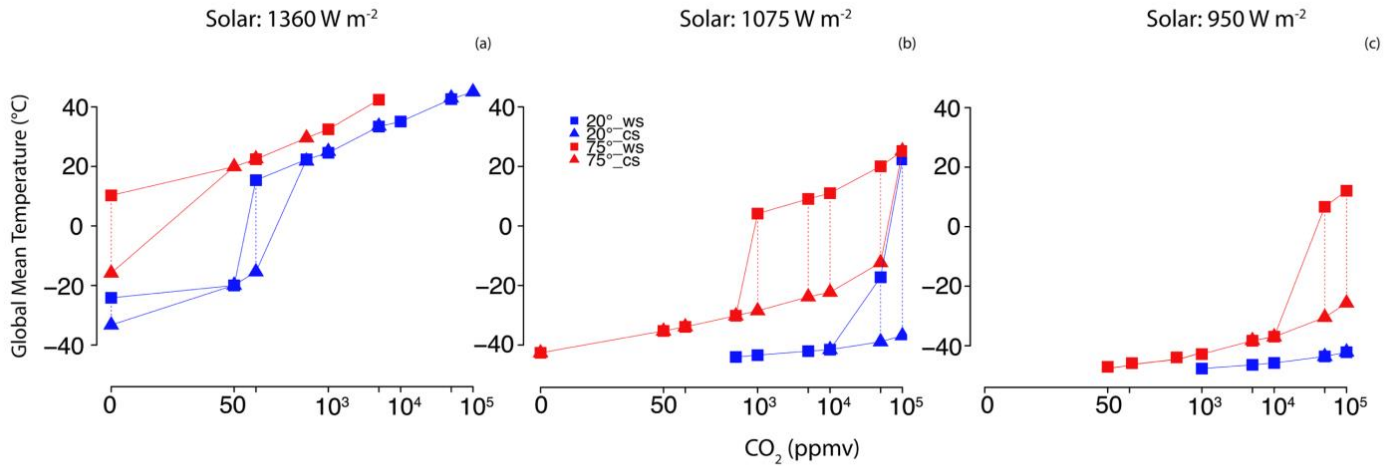

Figure 3: Global-mean temperature vs. $CO_2$ concentration for $S_0$=1360 (left), $S_0$=1075 (center), and $S_0$=950 (right). Blue and red data points are for 20° and 75° obliquity, respectively. Squares represent simulations with a warm start initial condition, and triangles with a cold start (snowball) initial condition. Data points are connected by solid lines as possible climate trajectories as $CO_2$ increases, and bi-stables states are connected by vertical dashed lines.

We note that in the ROCKE-3D aquaplanets explored here, the only 75° planet at $S_0$=1360 with any sea ice is the zero $CO_2$ experiment initiated in a cold start configuration; at 20° obliquity, sea ice vanishes beginning with the 1000 ppm $CO_2$ runs, regardless of initial conditions. The only full snowball state at $S_0$=1360 is the zero $CO_2$ experiment initialized in a cold start.

At $S_0$=1075, nearly all 20° obliquity planets feature global-mean temperatures below the freezing point (Figure 3b), except for the 10% $CO_2$ simulation initialized in a warm start. Of the rest, all are either in full snowball states (for runs with $CO_2$ ≤ 10,000 ppm, and at 5% or 10% $CO_2$ for the simulations initialized in a cold start) or have a stable equatorial waterbelt (5% initialized in a warm start). This results in a temperature jump from -39°C in the cold start configuration compared to -17°C in the warm start, shown in Figure 4b for <20°, 5% $CO_2$>.

For the same $S_0$=1075, all 75° obliquity warm start planets are above the freezing point (and ice-free everywhere) for $CO_2$ concentrations as low as 1,000 ppm. There is rich structure of multiple stable states in these experiments from 1,000 ppm-5% $CO_2$, with sea ice forming in the cold start experiments up until 5% $CO_2$. In no case however, are these complete snowballs (Figure 4-6).

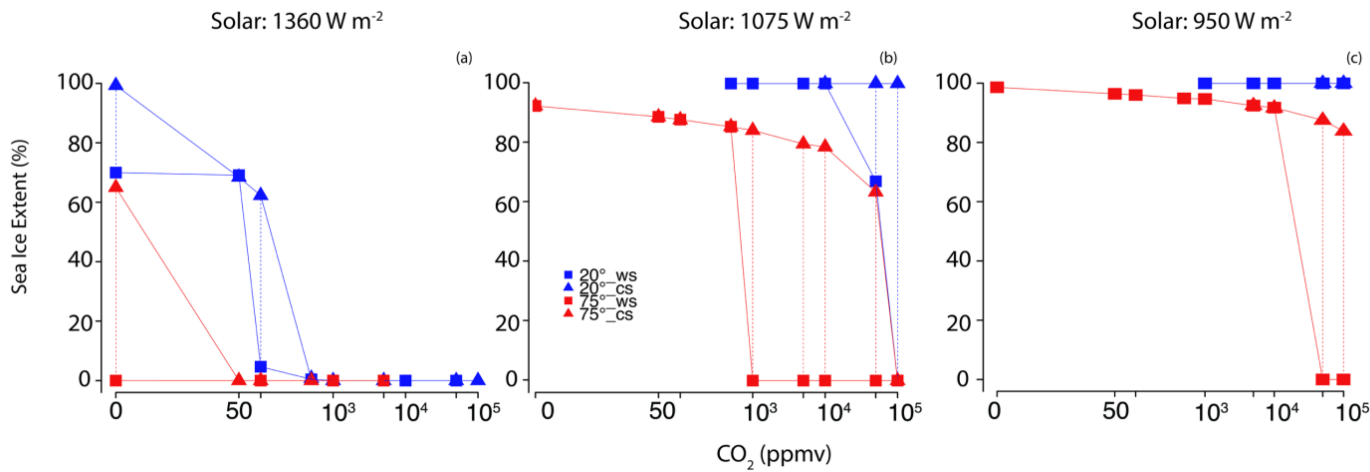

Figure 4: (a-c) As in Figure 3, except for sea ice cover.

At $S_0$=950, all 20° obliquity planets are in complete snowballs. At 75°, there is a small amount of polar ice melting during the summer for all $CO_2$ concentrations and a broader swath of polar liquid water opens up at higher $CO_2$ concentrations; by 10,000 ppm of $CO_2$, approximately 25% of the planet experiences some amount of sea ice melt throughout the year. For the warm start simulations at 5 and 10% $CO_2$, the planet is ice-free.

Figures 5 and 6 show example sea ice coverage (for the annual and JJA mean, respectively) selected from various runs in the suite of simulations. These examples encapsulate the cryosphere regimes permitted in ROCKE-3D within the parameter space explored. At low obliquity, the planet either is in a complete open ocean state, a snowball, contains a moderate amount of sea ice in the polar regions (similar to Earth), or near-planetary coverage of sea ice with an equatorial waterbelt. At high obliquity, meridionally confined equatorial ice-belts can develop, but are only observed in cold start experiments (see also Kilic et al., 2018). However, in all high obliquity experiments considered, liquid water exists in the polar regions at some point during the year. In many cases, open ocean persists in the winter, likely due to residual ocean heat buildup from the warm season (Ferreria et al., 2014). At $S_0$=950, nearly complete snowball conditions only develop at $CO_2$ concentrations as low as 1,000 ppm.

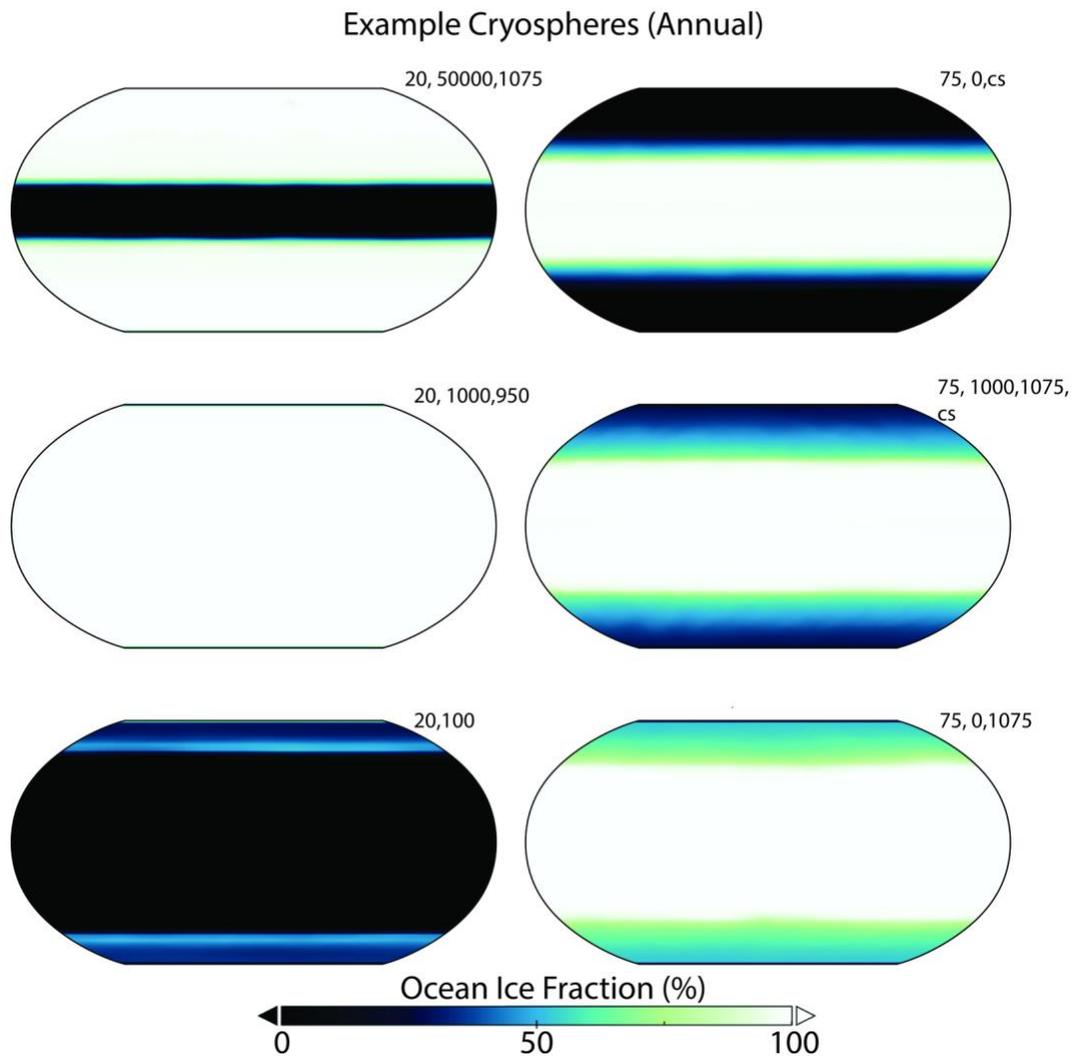

Figure 5: Select cases of annual-mean sea ice fraction (black = open ocean, white = ice-covered) from the suite of simulations performed. If labelled on upper-right, the first number indicates obliquity, second number is $CO_2$ in ppm, the third number is stellar constant in W m$^{-2}$, and "cs" denotes a cold start run.

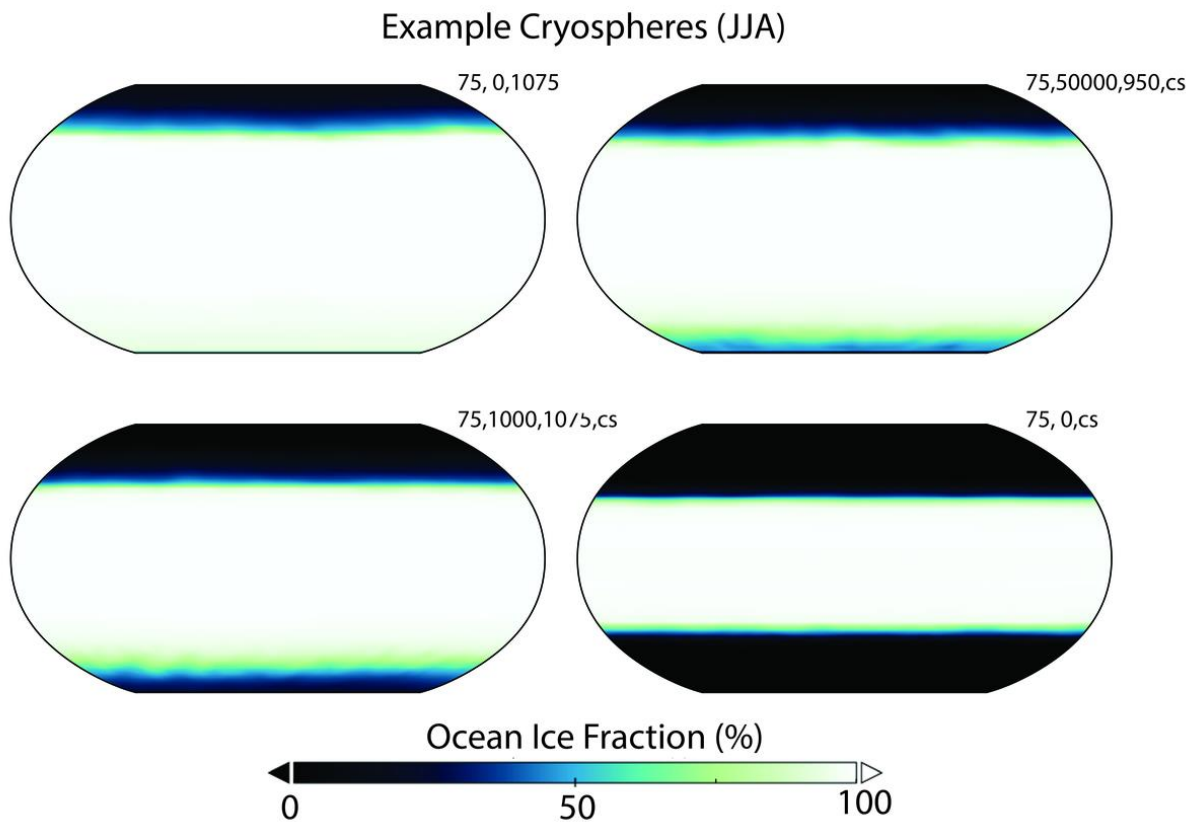

Figure 6: As in Figure 5, except for the JJA season.

High obliquity climates are highly resistant to a hard snowball state even for cases in which global-mean temperatures are well below freezing. Over the global-mean temperature range of -40° to 0°C, there is generally less sea ice for a given global temperature than at low obliquity (Figure 7a) due to the intense stellar heating and seasonality at high latitudes. We highlight global-mean temperatures below the freezing point in Figure 7a since there is zero or very small amounts of sea ice for warmer global temperatures and thus minimal habitability implications. For global-mean temperatures colder than about -25 °C, all of the low obliquity planets are in full snowball states, while the 75° obliquity experiments retain a small fraction of unfrozen area near the poles. For the moderately cold temperatures over this range (-25 °C to -10°C), the cryospheres generally exist as equatorial waterbelt states (for low obliquity) or equatorial ice-belts (for high obliquity), resulting in comparable areal coverage of ice and open ocean.

In Figure 7b, we show the fraction of planet that is either temporarily or permanently habitable for each experiment at $S_0$=1360, 1075, or 950. Permanently habitable conditions occur at a location if ocean ice fraction does not exceed 15% in the grid cell for any month of the year. Temporarily habitable conditions occur in a grid cell if ocean ice fraction does not exceed 15% for at least one month of the year, but less than all 12 months. On this plot, we note that a large number of experiments are in hard snowball states (zero permanently or temporarily habitable area, n =15) or in completely ice-free states (zero temporarily habitable area and 100% permanently habitable area, n =25).

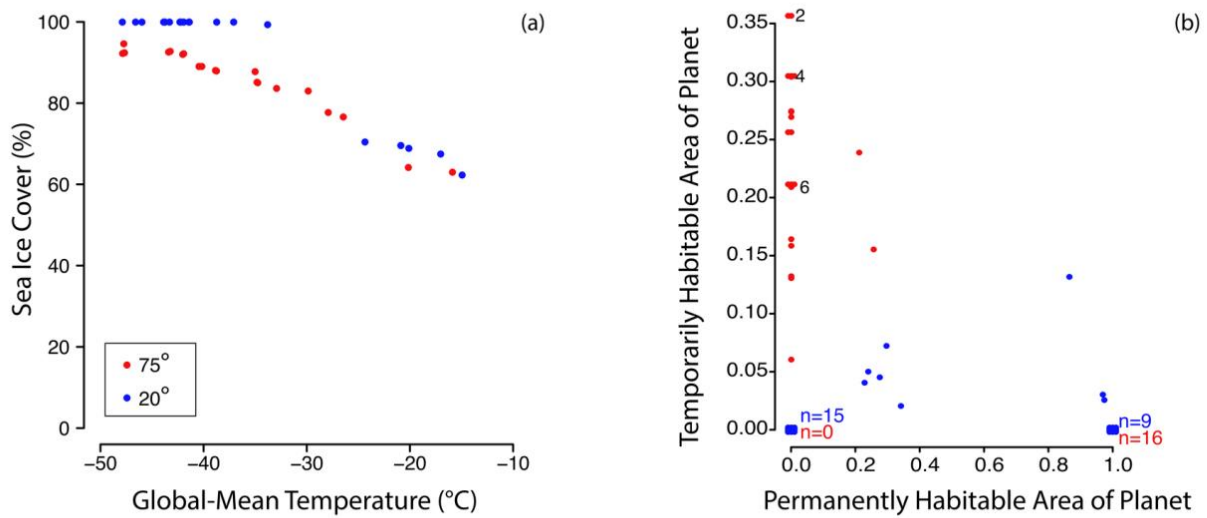

Figure 7: (a) Global sea ice cover plotted vs. global-mean temperature for all experiments at $S_0$=1360,1075, or 950 in which global temperatures are below freezing. Note that no experiment equilibrates between 0 and -10°C. (b) Temporarily habitable area vs. permanently habitable area (as defined in section 3.1) for all experiments at $S_0$=1360,1075, or 950. For both panels we show all data points in the suite of simulations as circles for 20° (blue) or 75° obliquity (red). Numbers (n=) are shown in Figure 7b for the number of simulations with zero temporarily habitable conditions and either 0% or 100% permanently habitable conditions, due to large data overlap.

Of the 20° obliquity experiments in which the planet is not either 0% or 100% permanently habitable, most have a moderate (20-40%) permanently habitable area and a very small (<10%) temporarily habitable area. These are equatorial waterbelt climates with marginal amounts of melting near the sea-ice edge during the annual cycle. Other experiments (e.g., $S_0=1360$ with 100 ppm $CO_2$) are in a regime similar to Earth with primarily open ocean conditions but with small amounts of polar sea ice. The seasonal cycle in sea ice formation and melting results in a small amount of temporarily habitable area (shown toward the bottom-right of Figure 8b with conditions of >80% permanently habitable area and less than 15% of temporarily habitable area).

At 75° obliquity, of the experiments that are not completely ice-free, a greater population of data points reside toward the upper-left portion of the figure where there is a small fraction of permanently habitable area and a relatively large fraction of temporarily habitable area. None of the experiments considered at high obliquity have both zero temporarily or permanent habitable surface area (i.e., complete freezing year-round), supporting the suggestion that it is very difficult for hard snowball conditions to develop on high obliquity bodies. However, for cold planets at 75°, most exhibit an equatorial ice-belt or nearly complete snowball conditions with seasonal sea ice formation and melting near the poles. This results in large fractions of the planet (>20% in many experiments) in a temporarily habitable regime, a rare occurrence for low obliquity experiments.

The $S_0=950$ experiment corresponds to an orbital distance of ~1.2 AU around the Sun, at which all of our 20° experiments would require more $CO_2$ (or higher atmospheric pressure or the presence of other absorbers) than we explored to initiate deglaciation, while all of the 75° simulations exhibit at least some amount of liquid water during part of the year. This suggests that a lower atmospheric $CO_2$ inventory may be expected on high obliquity planets with an active weathering feedback, for a given orbital distance. A calculation of weathering rates on high obliquity bodies for various stellar fluxes and topographies is left for future work.

3.2. Enhanced Global Temperatures at High Obliquity

Not surprisingly, high obliquity planets in the $S_o=1360$ experiments have lower planetary albedos (Figure 8a bottom, shown for $S_o=1360$) than their 20° counterparts that have some climatological sea ice. When both high and low obliquity planets are ice-free at $S_o=1360$, the high obliquity versions still feature a lower planetary albedo by 1-2% for $CO_2$ abundances from 500 ppm to 1000 ppm. One exception is the 75° case at 5000 ppm, that is brighter (~2% increase in global, annual-mean albedo) than at 20°, although in this experiment annual global temperatures are still warmer by 9°C in the high obliquity case.

The tendency for reduced planetary albedo on high obliquity worlds without an ice-albedo feedback involves an interaction between seasonal heating and cloud cover on the sunlit parts of the planet. Unlike Earth, storm tracks and baroclinic activity occur only in the summer hemisphere on high obliquity worlds (Ferreira et al., 2014), while the winter hemisphere is in darkness and heated nearly uniformly only by ocean heat fluxes into the atmosphere. Moist convection occurs in the winter hemisphere (not shown), although the peaks in low cloud cover, and the condensed water mixing ratio, shift into the summer hemisphere as stellar heating migrates toward the pole. Kang (2019) noted that due to the ocean's heat capacity, the summer mid-to-high latitude surface temperature increase and development of low cloud cover lag the peak solar heating by 1-2 months (see our Figure 9 from ROCKE-3D results), an effect that results in enhanced solar absorption relative to a low obliquity planet. Thus, our own results also support the claim that a delayed development of cloud cover in response to strong seasonal heating operates as a mechanism for lower planetary albedo and relative warmth at high obliquity.

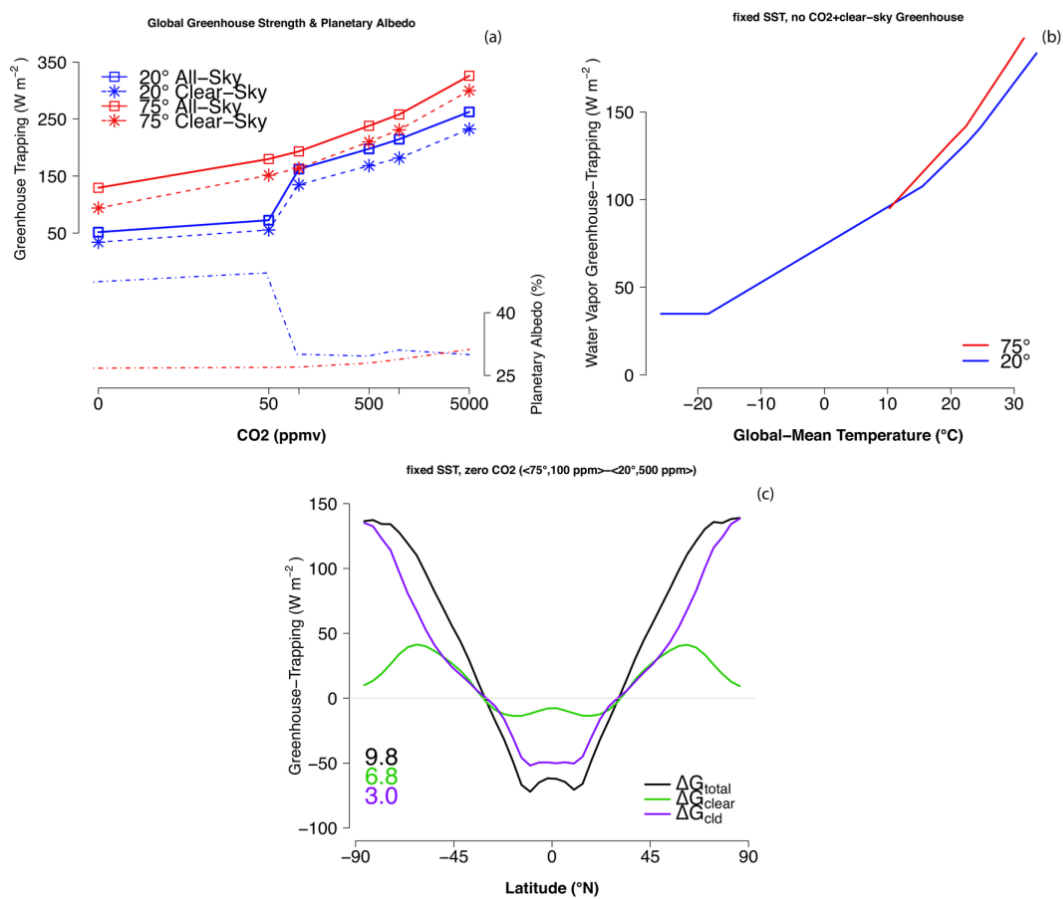

Figure 8: (a) (Top of plot) Global greenhouse-trapping (in W m$^{-2}$) for clear-sky and all-sky conditions for the $S_0$=1360 experiments from zero to 5,000 ppm of $CO_2$ at 20° (blue) and 75° (red) obliquity. (Bottom of plot, dot-dashed) Planetary albedo for the same experiments. (b) Clear-sky greenhouse effect in fixed SST, zero $CO_2$

experiments (described in main text) vs. global-mean temperature (c) All-sky change in greenhouse effect vs. latitude for the fixed SST simulations with zero CO2 based on the prescribed conditions for the coupled <75°,100 ppm, 1360 W m$^{-2}$> minus <20°,500 ppm, 1360 W m$^{-2}$> experiments (black) and decomposed into clear-sky (green) and cloudy sky (purple) contributions. Global-mean values shown in numbers on bottom-left of panel (c).

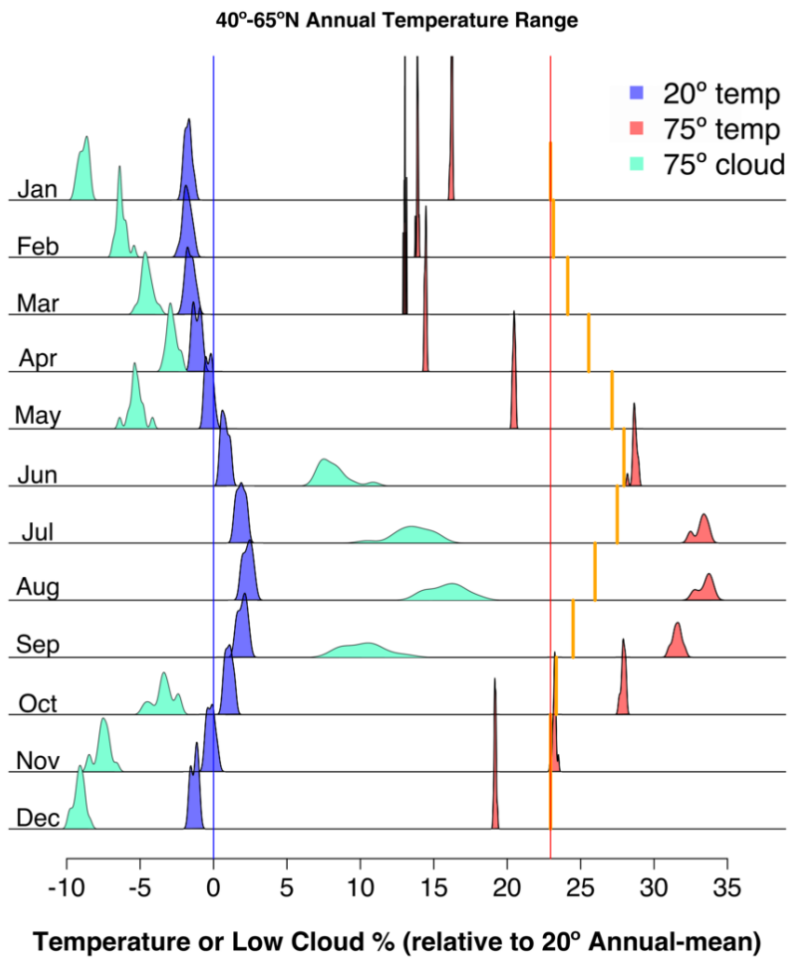

Figure 9: Monthly evolution of 40-65°N average temperatures ($S_0$=1350 and 500 ppm $CO_2$) for 20° (blue) and 75° obliquity (red), each expressed as anomalies relative to the 20° annual-mean. Low cloud cover anomalies (%) for the 75° obliquity simulation only is shown in aquamarine. Distributions reflect in-month variability taken over 30 years at the end of the simulation period. Solid orange lines show the monthly evolution of 40-65°N stellar flux, normalized and offset.

Another factor that likely contributes to enhanced global temperature is the re-distribution of moisture from low to high latitudes. Although the greenhouse effect of water vapor is often thought of as a passive amplifying feedback to the global temperature (via the saturation vapor pressure constraint imposed by the Clausius-Clapeyron relation), it may also serve as the cause of the enhanced warmth since the dynamical regime prevailing on high obliquity bodies is markedly different than at low obliquity and the spatio-temporal distribution of water vapor and the degree of atmospheric sub-saturation play a critical role in the planetary energy budget (Pierrehumbert et al., 2007). There need not be a unique relationship between global-mean temperature and water vapor greenhouse trapping. We note that the intense heating of high latitudes for part of the year, and the collapse of intense descent in subtropical subsiding regions associated with the vanishing of a strong annual-mean Hadley circulation (Kang et al., 2019) results in a smoother moisture field than on Earth and higher relative humidities aloft over much of the planet (Figure 10).

In Figure 10a, we show the precipitable water (the mass-weighted column integral of specific humidity) for the 20° and 75° obliquity planets at 500 ppm and $S_0$=1360. At high obliquity, the high latitudes contain more moisture than low latitudes, and with a pronounced seasonal cycle. The annual and zonal-mean circulation at high obliquity (Figures 10 c and d, respectively) promotes a less developed intertropical convergence zone and cloud cover (not shown) compared to low obliquity, and subtropical regions become cloudier and contain more moisture. We also compare the fractional change in moisture at <75°, 100 ppm> and <20°, 500 ppm> relative to the <20°, 100 ppm> experiment. By coincidence, the <75°, 100 ppm> and <20°, 500 ppm> have virtually identical global-mean temperatures (the high obliquity case is ~0.05°C warmer) and similar amounts of water vapor, although the distribution of water vapor is markedly different.

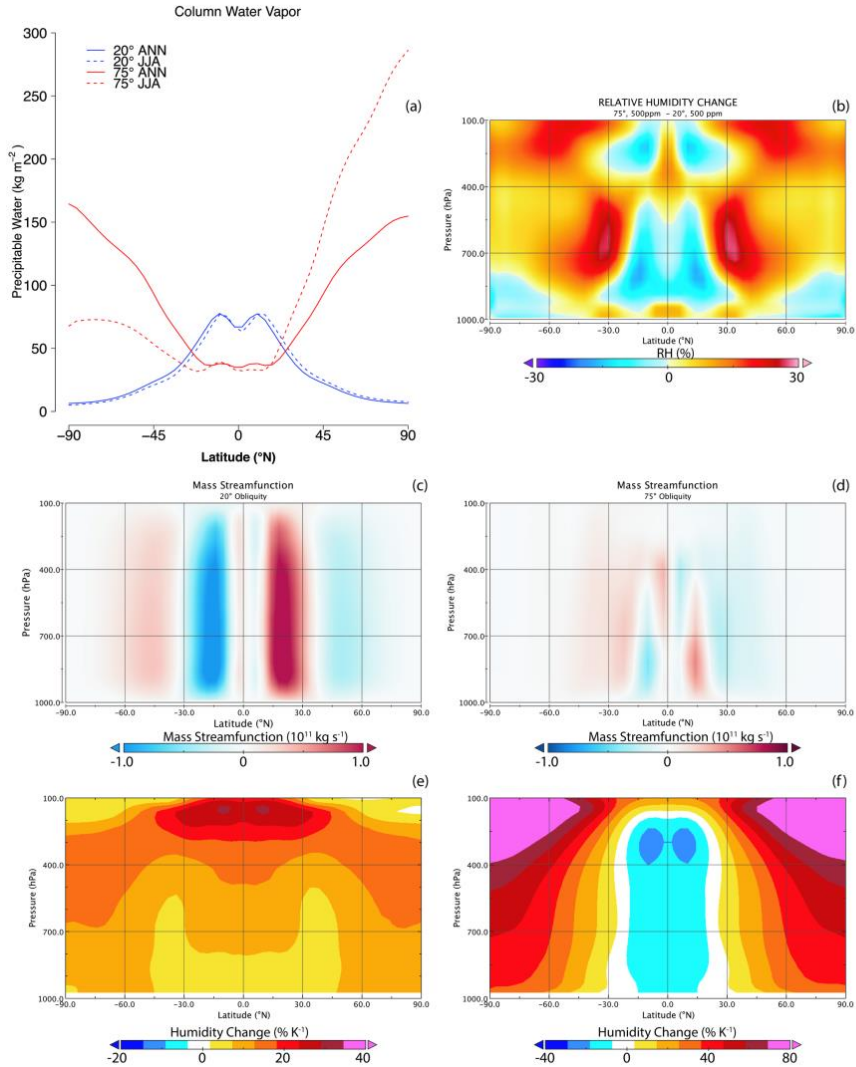

Figure 10: (a) Precipitable water (kg m$^{-2}$) vs. latitude (annual-mean in solid, JJA in dashed). (b) Relative Humidity (%) difference between the 75° and 20° experiments. (c and d) Mass streamfunction ($10^{11}$ kg s$^{-1}$) at 20° and 75°, respectively. Plot (a-d) use the $S_0$=1360 simulations at 500 ppm CO$_2$ (e) Fractional changes in specific humidity [d(log q)/(q dTs)] between the <20°,500 ppm, 1360 W m$^{-2}$> and <20°,100 ppm, 1360 W m$^{-2}$> experiments. (e) Fractional change in specific humidity between the <75°,100 ppm, 1360 W m$^{-2}$> and <20°,100 ppm, 1360 W m$^{-2}$> experiments.

.

Figure 8a shows the all-sky and clear-sky greenhouse trapping in the warm-start experiments for $S_0$=1360 only. Greenhouse trapping is defined (in W m$^{-2}$) as the difference between upwelling thermal radiation at the surface ($F_s$) and that at the TOA ($F_t$). Clear-sky calculations are performed with an offline second pass through the radiation code in which clouds are transparent to radiative fluxes but in which the climate is not affected.

A direct comparison of greenhouse trapping in the coupled experiments (Figure 8a) shows a monotonic increase in greenhouse trapping with increasing $CO_2$ for both 20° and 75° experiments, with the high obliquity version of each experiment systematically higher than the low obliquity case. The increase in greenhouse trapping includes the direct influence of $CO_2$ and the varying amount of water vapor arising for thermodynamic reasons. Because the high obliquity experiments are warmer than their low obliquity counterparts, they also have a larger moisture content.

To gain insight into the significance of the spatial structure of water vapor on clear-sky greenhouse trapping on low vs. high obliquity planets, we performed an additional set of fixed sea surface temperature (SST) simulations in which $CO_2$ is zeroed out in order to remove the varying influence of the non-condensing greenhouse effect. The prescribed SST and sea ice fields were taken from each of the warm-start coupled simulations at $S_0$=1360. The resulting climates and water vapor inventory are virtually identical to that of the corresponding coupled simulation (not shown) allowing for a comparison of the $CO_2$-removed greenhouse effect between the simulations. Clear-sky greenhouse trapping is plotted against global-mean surface temperature (Figure 8b), with the goal of identifying whether for a given temperature, clear-sky global greenhouse trapping is higher for high or low obliquity planets. As shown in Figure 8b, when normalized against global-mean temperature, the $CO_2$-removed greenhouse effect is still higher in the 75° obliquity experiments. We interpret the differences as arising from differences in the spatial and temporal distribution of water vapor at high obliquity, and a causal explanation for the higher global-mean temperatures that help maintain temperate conditions approaching the OHZ.

To further support the argument that high obliquity planets at a given global-mean temperature will tend to have a stronger water vapor greenhouse effect, in addition to showing the water vapor greenhouse trapping across all of the warm start simulations for $S_0$=1360, Figure 8c shows the difference in the latitudinal profile of greenhouse trapping in the no-$CO_2$ fixed SST simulations between the <75°, 100 ppm> case and the <20°, 500 ppm> case, that as mentioned before, have approximately the same global-mean temperature. Contributions from clear-sky and cloudy-sky contributions are calculated using the method in Rencurrel and Rose (2018, see their appendix C) in which:

$$\Delta G_t = \Delta G_t^{clear} + \Delta G_t^{cloud} \quad (2)$$

$$\Delta G_t^{clear} = (1-c)\big[\Delta F_s - \Delta F_{t,clear}\big] \quad (3)$$

where $\Delta G_t$ is the difference in total-greenhouse effect between the <75°, 100 ppm> and <20°, 500 ppm> experiments, c is total cloud cover taken from the low obliquity case, and $\Delta F_s - \Delta F_{t,clear}$ is the difference between the clear-sky greenhouse trapping in the two experiments. $= \Delta G_t^{clear}$ is the contribution of clear-sky greenhouse processes to the difference in climate between the two experiments. $\Delta G_t^{cloud}$ is the cloud effect, calculated as a residual.

The total greenhouse trapping is ~10 W m$^{-2}$ higher in the high obliquity case, of which $= \Delta G_t^{clear}$ contributes approximately 7 W m$^{-2}$ in the global-mean. The cloud contribution has a spatial structure that closely tracks the total greenhouse trapping, but is positive in the subtropics and at high latitudes and substantially negative at low latitudes, due to less low latitude high cloud cover generated by intense moist convection, resulting in a smaller global-mean value.
These results indicate that both clouds (via longwave trapping and the influence of reduced planetary albedo) and the water vapor distribution contribute to global warming in high obliquity climates. We suggest based on Figure 8b and 8c that the varied moisture field on high obliquity planets can be interpreted as a cause for warmer temperatures across the set of experiments.

3.3. Effect of Ocean Heat Transport

We have conducted a parallel set of select simulations described above except with a 50 m thermodynamic slab ocean (with zero horizontal and vertical heat transport and lower heat capacity than our deeper dynamic ocean) to explore the significance of ocean dynamics and heat storage. Seasonality is comparable to that of the coupled simulations, indicating a significant role for ocean heat storage (Ferreira et al., 2014) rather than horizontal ocean heat transport. We note that the increase in global temperature with obliquity and the presence of bi-stability remain robust in the absence of ocean heat transport, although the location of the bifurcation points and mean sea ice cover are strongly sensitive to the ocean configuration (Figure 11). For example, at $S_o$=1360 and 20° obliquity, nearly half the planet is covered with sea-ice at 500 ppm of $CO_2$ (compared to ice-free conditions in the coupled experiment), and there is moderate sea ice at high latitudes up to 1,000 ppm of $CO_2$ (Figure 11 compared to Figure 4). At 75° obliquity, there is equatorial sea ice in the 50 ppm $CO_2$ experiment initialized in a cold start, but this same configuration is ice-free with the dynamic ocean. At $S_o$=1075 and 1,000 ppm $CO_2$, the warm start 75° planet is ice-free in the coupled experiments, but is mostly glaciated in the q-flux version. Similarly, the 20° warm-start planet is ice-free at $S_o$=1075 and 10% $CO_2$, but is in an equatorial waterbelt state in the q-flux experiment. Furthermore, for conditions where an equatorial ice-belt (waterbelt) exists at high (low) obliquity, the edge of the sea ice extent can be noticeably affected. However, once sea ice vanishes in both the coupled and q-flux version, the global-mean temperature difference between the two is at most several degrees (not shown).

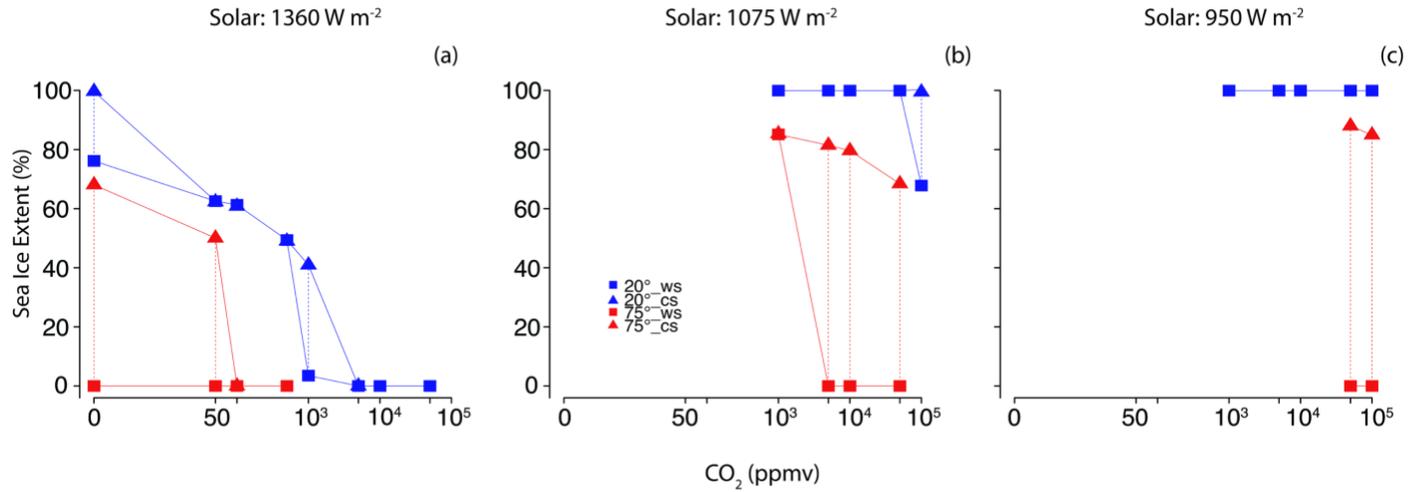

Figure 11: Global-mean sea ice extent vs. $CO_2$ concentration for $S_0$=1360 (left), $S_0$=1075 (center), and $S_0$=950 (right) in the slab-ocean (zero horizontal ocean heat transport) simulations. Blue and red data points are for 20° and 75° obliquity, respectively. Squares represent simulations with a warm start initial condition, and triangles with a cold start (snowball) initial condition. Compare to the coupled experiments in Figure 4.

3.4. Multi-bar Atmospheres

Figure 12 shows results from the 5 bar $CO_2/CH_4/N_2$ and 10 bar $H_2/N_2$ experiments described in section 2. In the 5 bar atmosphere (Figure 11a), both 20° and 75° planets move from snowball to ice-free conditions between $S_0$=500 and $S_0$=600. At $S_0$=550, the 75° simulation has undergone deglaciation while the 20° planet remains in a snowball. Due to the inverse square law, the separation in orbital distance between these two limits is fairly small (~0.15 AU around the Sun), indicating a relatively small window for which obliquity may substantially affect the glacial state of a dense atmosphere planet at the OHZ. At $S_0$=600, the 20° case has a modern Earth-like climate that is ~14°C near the equator and approaches the freezing point at the poles, while the high obliquity planet is 8.2°C warmer in the global mean. Thus, high obliquity has less impact on habitability at the OHZ when multi-bar $CO_2$ atmospheres are expected to develop, although it can tip the scale toward warmer climates within a certain part of parameter space. In the 10 bar experiments (Figure 12b), increased obliquity generates no change in the fractional habitability for threshold cases (deglaciation begins $S_0$=125-150 for both obliquity regimes) and global-mean temperatures remain within 1°C of each other for all four stellar flux values considered.

The effect of the longer orbital period is shown in Figure 12c where JJA temperatures are plotted for the 75° experiments in the 5 bar atmosphere. A separate set of simulations were run for $S_0$=500, 550, and 600 in which the orbital period is kept at Earth's value. For global temperatures above the freezing point, the effect of the orbital period is very small, but for snowball conditions the effect can be large due to the small heat capacity of the sea ice covered surface and longer seasonal length to respond to the incident stellar flux. We note that for the 10 bar experiments, seasonality is minimal at high obliquity (not shown). The difference in temperatures between the two poles is only ~5°C during JJA in the climates above freezing and ~10°C in the $S_0$=125 experiment that is in a snowball.

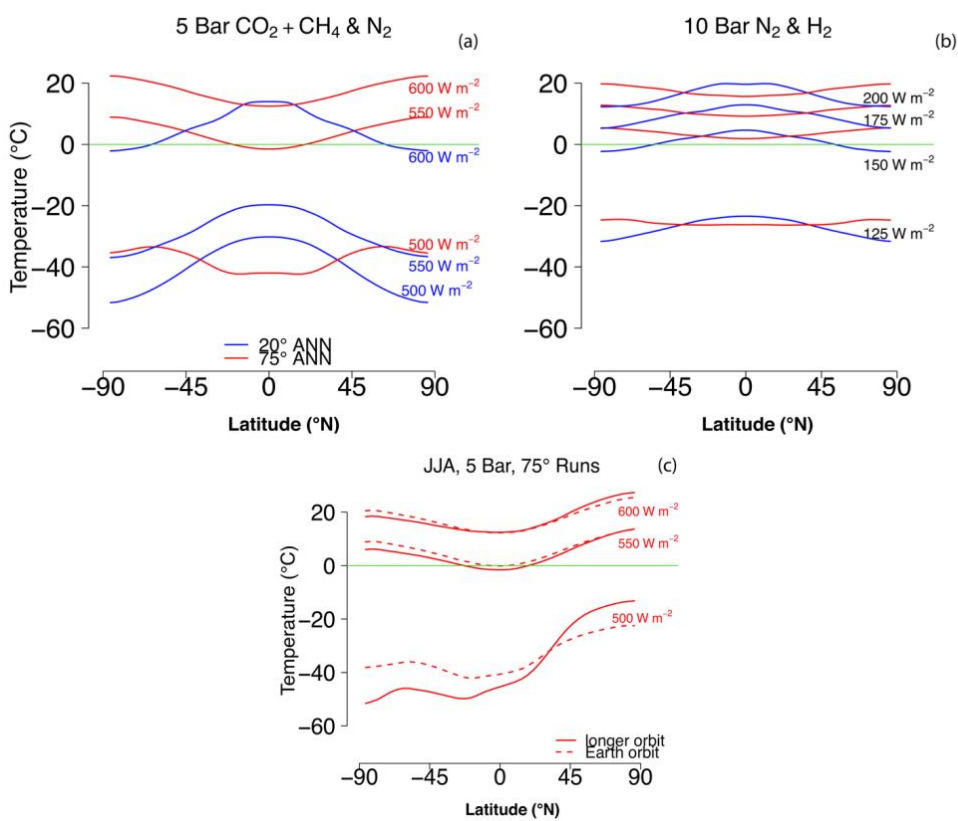

Figure 12: (a) Latitudinal temperature profiles (annual) for 5 bar $CO_2/CH_4/N_2$ simulations (described in text) for 20° (blue) and 75° (red) obliquity, with stellar constants labelled from 500-600 W m$^{-2}$. (b) For 10 bar $H_2/N_2$ with stellar constants

labelled from 125-200 W m$^{-2}$. (c) JJA temperature for 75° obliquity simulations only, in simulations with elongated orbital period (solid) and Earth's orbital period (dashed).

These results are consistent with arguments based on energy balance models (Williams and Kasting, 1997) that the impact of obliquity on habitability should diminish in dense atmospheres. However, we suggest that although the long radiative time constant in such atmospheres mutes the extreme seasonal cycle of high obliquity planets, this actually helps hinder fractional habitability by removing the prospect of localized summer melt. The same negative feedback that increases the greenhouse effect to offset low stellar fluxes becomes the bottleneck that limits local reservoirs of polar water to persist in a glacial climate.

## 4. CONCLUSIONS

We used the NASA ROCKE-3D GCM to probe the effect of increased obliquity for a 1 bar atmosphere at three different orbital distances gradually approaching the OHZ and for varying $CO_2$ concentrations. It is shown that global temperatures increase at high obliquity across the entire parameter space explored. Furthermore, liquid water persists in polar regions for global-mean temperatures approaching values as cold as -40°C, due to intense summer heating at the poles. This suggests that high obliquity planets may be favorable candidates for habitability on planets with weaker incident stellar flux than Earth. However, the effect on fractional habitability diminishes near the OHZ beyond the point where dense atmospheres would be expected to accumulate on planets with active weathering cycles.

Global-mean temperature increases at high obliquity, due to ice-albedo feedbacks (in cold climates) or the water vapor and cloud responses (in temperate and hot climates) associated with the different dynamical regimes on these bodies. This is not interpreted as entirely due to a feedback to global-mean temperature, as is appropriate in the limit of small thermodynamic changes in the context of contemporary climate change (Held and Soden, 2000). Rather, the development of a different spatial temperature structure and lack of strong dry zones (either due to very cold temperatures or those associated with intense descending motion) as stellar heating moves from pole-to-pole results in a stronger global-mean water vapor greenhouse effect for a given global-mean surface temperature. Clouds also play a role by responding to the incident heating that occurs in the summer, and the development of large-scale cloud cover lags the peak heating (Kang, 2019). However, the cloud feedbacks should be tested across other GCMs to explore the parameter space in which high obliquity reduces planetary albedo.

High obliquity planets, particularly those that did not start early in their history in a snowball state, are highly resilient to global or even wintertime glaciation. Bi-stability is present at both high and low obliquity in ROCKE-3D, with

24the possibility of equatorial ice-belts and varying amounts of stable liquid water in the polar regions at low stellar fluxes.

We suggest that for planets with a moderately weaker incident stellar flux relative to Earth, high obliquity generates more favorable conditions for habitability, either locally or globally. Furthermore, these planets require very low $CO_2$ in order to sustain hothouse climates even beyond Earth's orbit. Seasonality may be a hindrance for various topographic regimes, especially configurations with polar-centric continents since open ocean can no longer mute very large temperature excursions except near the coasts.

## ACKNOWLEDGMENTS

CC's research was supported by an appointment to the NASA Postdoctoral Program at the NASA Goddard Institute for Space Studies, administered by the Universities Space Research Association under contract with NASA. AD and MW were supported by the NASA Astrobiology Program through collaborations arising from our participation in the Nexus for Exoplanet System Science (NExSS). Igor Aleinov and Reto Ruedy provided support for running the model experiments described in the paper. Computing resources for this work were provided by the NASA High-End Computing (HEC) Program through the NASA Center for Climate Simulation (NCCS) at Goddard Space Flight Center.## REFERENCES

Armstrong, J. C., Barnes, R., Domagal-Goldman, S. et al. 2014, AsBio, 14, 277

Bin, J., Tian, F., & Liu, L. 2018, EPSL, 492, 121

Bitz, C. M., Holland, M. M., Hunke, E. C., & Moritz, R. E. 2005, JClim, 18, 2903

Bolmont, E., Libert, A. S., Leconte, J., & Selsis, F. 2016, A&A, 591, A106

Borucki, W. J., Koch, D. G., Basri, G., et al. 2011, ApJ, 728, 117

Boutle, I.A., N.J. Mayne, B. Drummond, et al., 2017, A&A, 601, A120

Claire, M. W., Sheets, J., Cohen, M. et al. 2012, ApJ, 757, 95